\documentclass[aps,prd, twocolumn,superscriptaddress,groupedaddress]{revtex4}
\usepackage {graphicx}  
\usepackage {dcolumn}   
\usepackage {bm}        
\usepackage {amssymb}   
\usepackage {color}
\usepackage {amsmath}
\usepackage {amsfonts}
\usepackage {amsthm}
\usepackage {mathrsfs}
\usepackage {natbib}
\usepackage {latexsym}
\usepackage {dsfont}
\usepackage {txfonts}
\usepackage {rotating}
\usepackage {multirow}
\usepackage {hhline}
\usepackage {hyperref}
\usepackage {bm}
\usepackage {appendix}
\usepackage {url}
\usepackage {acronym}
\usepackage {enumitem}
\usepackage {aas_macros}
\usepackage{ulem}
\usepackage{braket}
\hypersetup{
  colorlinks=true,        
  linkcolor=black,         
  citecolor=cyan,         
}

\begin{document}

\title{\Large \textbf{Gravitons probing from stochastic gravitational waves background}}

\author{Hongguang Zhang}
\affiliation{School of Physics and Technology, Wuhan University, Wuhan 430072, China}

\author{Xilong Fan}
\email[Xilong Fan:]{xilong.fan@whu.edu.cn}
 \affiliation{School of Physics and Technology, Wuhan University, Wuhan 430072, China}

\author{Yihui Lai}\email{yhlai@umd.edu}
\address{Physics department, University of Maryland, College Park, Maryland, 20740, USA}

\begin{abstract}
Quantum gravity is a challenge in physics, and the existence of graviton is the prime question at present.  We study the detectability of the quantum noise induced by gravitons in this letter. The correlation of the quantum noise in the squeezed state is calculated, but we are surprised to find that the result is the same as those from Grishchuk et.al. and Parikh et.al. This implies that they are essentially about one thing: the quantum noise from the gravitons. The further discussion shows the quantum noise should be completed from the leading order of the interaction between gravitation and the detector. The squeezed factor is estimated, and with it, the spectrum of the quantum noise is found to be of the form  $\Omega_0 \sim \nu^{4+a_0 \beta+b_0}$. After comparing it to the sensitivities of several gravitational wave detectors, we conclude that the quantum noise is detectable in the future.

\end{abstract}


\maketitle

 
\section{Introduction}

Quantum gravity is one of the most intensively studied areas in physics. Theories like superstring and quantum loop are approaches to this direction. However, neither one is satisfactory. Some physicists think that gravity may not be quantized canonically \cite{jacobson}. If so, as a clear consequence, the existence of graviton, a massless helicity-two particle of gravitational unity in quantum mechanics \cite{feyman1, feyman2,feyman3, feyman4}, is questionable.

Dyson \cite{dyson} discussed the possibility to detect a single graviton, and he found there is no conceivable experiment in our universe for that since we have to increase detector sensitivity by some 37 orders of magnitude. The detection of primordial gravitational waves implies a discovery of gravitons, however, the usual detection of the gravitons is impossible with current experimental techniques \cite{Kanno2018, Kanno2019}. One possibility is to utilize the unique features of the squeezed primordial gravitational waves from inflation, as it is Gaussian but non-stationary. This can be used to distinguish the background created from inflation as well as other types of stochastic background. Unfortunately, Allen et. al show that is still practically impossible in \cite{allen}, as the conceived experiment would have to last approximately the age of the Universe at the time of measurement. The other possibility is to probe the gravitons in squeezed state through its statistical properties in \cite{grishchuk1,grishchuk2,grishchuk3}, which seems to be some optimistic, though one doesn't know how to distinguish the background from the early universe from others in experiment. Recently, an alternative method of probing the quantum coherent aspect of gravity from the updated quantum information techniques has been proposed. It offers an unambiguous and prominent witness of virtual gravitons \cite{vedral1, vedral2, vedral3}, but that needs to be scrutinized further.

More recently, Parikh, Wilczek, and Zahariade suggested a direction of probing the gravitons from the induced quantum noise in the arms of gravitational wave detectors, as the geodesic separation of a pair of freely falling masses includes a stochastic component in the presence of the quantized gravitational field \cite{wilczek1, wilczek2, wilczek3}. Quantum noise in the arms of gravitational wave detectors is produced by introducing the gravitons. Literature about the decoherence in effective theory in this direction can be found \cite{hu1,hu2,hu3,hu4,hu5,hu6,hu7,hu8}. The holographic version is in \cite{zurek}, and the equation of motion for the geodesic deviation between two particles from the Langevin-type equation is derived in \cite{soda}. We study the detectability of the quantum in this work: the correlation in squeezed state is calculated using the method shown in \cite{soda} but with a redefined quantum noise.

\section{The correlation of quantum noise} 

We start to describe the gravitational waves in Minkowski space, with the metric 
\begin{equation}
\label{metric}
ds^2=-dt^2 + (\delta_{ij} +h_{ij} dx^{i} dx^{j}),
\end{equation}
where $t$ is the time, and $x^{i}$ are the space, $\delta_{ij}$ and $h_{ij}$ are the Kronecker delta and the metric perturbation satisfying the transverse traceless conditions. The indices ($i,j$) run from 1 to 3. 

A system of gravitational waves interacting with a detector is considered in this work. The gravitational waves are canonically quantitized, and the quantum noise induced by gravitons is defined and studied. This system would lead to a loss of quantum decoherence, like the model shown in \cite{soda}. What we want to stress here is: our interaction picture is quite general, without needing a concrete action.

Denoting the index A to be the linear polarization modes $A= +, \times$, we quantitized gravitational wave field $\hat h^A_{\rm I}({\bf k},t)$ canonically, with the creation and annihilation operators in interaction picture as $\hat h^A_{\rm I}({\bf k},t)
      =\hat a_A({\bf k})u_k(t)+\hat a^\dag_A(-{\bf k})u^*_k(t)$.
The creation and annihilation operators satisfy the standard commutation relations $
\left[\,{\hat a}_A({\bf k})\,,\,\hat{a}^\dag_{A'}({\bf k'})\,\right]
=\delta_{{\bf k},{\bf{k}'}}\delta_{AA'}\,,
\left[\,{\hat a}_A({\bf k})\,,\,\hat{a}_{A'}({\bf k'})\,\right]=\left[\,{\hat a}^\dag_A({\bf k})\,,\,\hat{a}^\dag_{A'}({\bf k'})\,\right]=0$.
The wave function $u_k(t)$ is a mode function that can be properly normalized as $ \dot u_k(t)u_k^*(t)-u_k(t)\dot u_k^*(t)=-i$.
In Minkowski space, the vacuum $\ket{0}$ is defined by $\hat a_A({\bf k}) \ket{0} =0$, with the mode function is chosen as 
$u_k^{\rm M}(t)= e^{-ikt}/\sqrt{2k}$.
The mode function in squeezed state is given in terms of that in Minkowski space in (\cite{soda}) such as
\begin{equation}
\label{eq:squeezed}
u_k(t)^{sq}=  u_{k}^{M}(t) \cosh(r_k) -e^{-i\phi_k} u_{k}^{M*}(t) \sinh(r_k) ~,
\end{equation} 
where  $r_k$ and $\phi_k$ are the squeezed parameters. 

It is sure that we can divide the gravitational perturbation $\hat{h}_{I}^{A}$ around the classical Minkowski background with the graviton
\begin{equation}
\label{eq:gravition}
 \hat{h}_{I}^{A} = \delta \hat{h}_{I}^{A} + h_{cl}^{A} ,
\end{equation}
where $h_{cl}^{A}({\bf k},t) \!=\!\left<0|\hat{h}_{I}^{A}({\bf k},t)|0\right>$ is the classical background, and $\delta \hat{h}_{I}^{A}$ is the graviton in the presence of the classical background with $\left<0|\delta \hat{h}_{I}^{A}|0\right> \!=\!0$. In the quantitization, we define the finite volume as $V \!=\! L_x L_y L_z$ and the discretized k-mode as $k \!=\! (2\pi n_x/L_x, 2\pi n_y/L_y, 2\pi n_z/L_z)$,
where $n \!=\! (n_x, n_y, n_z)$ are integers. It is obvious that we can define the effective strain below as the quantum noise, as what is done in \cite{grishchuk1}, 
\begin{eqnarray}
\label{eq:noise}
\delta \hat{h}_{ij}(t) =\frac{\kappa}{\sqrt{V}}\! \sum_{\bf k, A} e^{A}_{ij}({\bf k} ) \delta \hat{h}^A_{\rm I}({\bf k},t)\ ,
\label{eq:noise}
\end{eqnarray}
where we take $\kappa \!=\!\!\sqrt{8\pi G}$ to make $\delta \hat{h}_{ij}(t)$ dimensionless. As discussed in \cite{grishchuk1, wilczek1, soda}, gravitons from the early universe might be in a squeezed-coherent state $\left|\xi, B\right>$ with
\begin{equation}
\left|\xi, B\right> = \hat{S}(\xi) \hat{D}(B)\left|0\right> ,
\end{equation}
where $\hat{S}(\zeta)$ and $\hat{D}(B)$ are the squeezing and the displacement operators. The operators are defined as
\begin{eqnarray}
\hat{S}(\zeta) \!\! &\equiv& \!\! \exp\left[\frac{1}{V}\! \sum_{\textbf{k},A}\left(\zeta^{*}_k \hat{a}_{A}(\textbf{k}) \hat{a}_{A}(\textbf{k}) \!+\!\zeta_k \hat{a}^{\dagger}(\textbf{k}) \hat{a}^{\dagger}_{A}(-\bf{k}) \!\right)\right] \!~, \\
\hat{D}(B)\!\! &\equiv& \!\! \exp\left[\frac{1}{V} \sum_{\textbf{k},A}\left(B_{k} \hat{a}^{\dagger}_{A}(\textbf{k}) - B^{*}_{k}\hat{a}_{A}(\textbf{k})\right)\right] ~,
\end{eqnarray}
where $\zeta  \equiv r_{k}\exp[i\phi]$ and $r_k$ are the squeezed parameter. $B_k$ is the coherent parameter.

Following the work of \cite{soda}, where a similar calculation with details is performed, we calculate the correlation in the squeezed state and give the result directly here
\begin{eqnarray}
\label{eq:correlation}
\left<\xi, B|\delta \hat{h}_{ij}(t) \delta \hat{h}_{kl}(t')|\xi, B\right> =\frac{4 \kappa^2}{\pi^2}\! \int\!\! \mathrm{~d}k~k^{2} Re[\mu_{k}^{s}(t)\mu_{k}^{s*}(t')] ~.
\end{eqnarray}
For large squeezing $r \gg 1$, the correlation becomes
\begin{eqnarray}
\label{eq:large}
\left<\xi,\! B|\delta \hat{h}(t)\delta \hat{h}(t')|\xi,\! B\right>\!=\!\frac{2\kappa^2}{\pi^{2}}\!\!\int\!\!\frac{dk}{k}k^2\!\left[\sin(\frac{\varphi}{2})sin(\frac{\varphi}{2}\!-\!k \delta t)\right]\!e^{2r_{k}}.
\end{eqnarray}
For $\delta t=0$ we have the following expression
\begin{eqnarray}
\label{eq:limit}
\left<\xi, B|\delta \hat{h}(t) \delta \hat{h}(t)|\xi, B\right>
\!=\!\int_{m_{ir}}^{m_{uv}}\!\frac{\mathrm{~d}k}{k} \delta h_{rms}^2 ~,
\end{eqnarray}
where the upper limit of the measurement $m_{uv}$ and the lower limit $m_{ir}$ are added, while the variance is defined as
\begin{eqnarray}
\label{rms1}
\delta h_{rms}^2\!&=&\!\frac{2 \kappa^2}{ \pi^{2}}  sin^2(\frac{\varphi}{2})k^{2}e^{2r_{k}} .
\end{eqnarray}
The phase factor $\varphi $ here should depend on $\nu$ in general, and it is the same as the non-stationary random process in \cite{allen}. The term of $sin(\frac{\varphi}{2})$ reflects the oscillatory behavior, which is a unique signature that could in principle be distinguishing regardless of the amplitude. However, as pointed out in \cite{allen}, this signature is not observable with a gravitational wave detector. Therefore, we will simply ignore this factor in the following calculation and focus on the amplification
\begin{equation}
\label{rms2}
\delta h_{rms}^2 \!=\! 8\kappa^2 \nu^{2}  e^{2r_{\nu}} ,
\end{equation}
where we define the squeezed factor $e^{r_{\nu}}$. For gravitational waves which are comfortably shorter than the Hubble radius, the amplitude of the graviton is amplified during the evolution of the universe. It is observed that the spectrum of energy density at present $\Omega_{0} (\nu)$ can be written as (see \cite{grishchuk1})
\begin{eqnarray}
\label{eq:omega}
\Omega_{0}(\nu) \!=\! \frac{\pi^2}{3} \delta h_{rms}^2  \left(\frac{\nu}{\nu_H}\right)^{2} \!=\! \frac{64\pi^3 G \nu^4}{3\nu_H^2}  e^{2r_{\nu}},
\end{eqnarray}

\section{Unify the picture}
We notice that the similar calculations are  done in literature: Grishchuk et.al give the strain of  gravitons in \cite{grishchuk1, grishchuk2,grishchuk3}, while Parikh et.al. give the power spectrum of the quantum noise in \cite{wilczek1, wilczek2, wilczek3}. With their results, we derive the corresponding energy density $\Omega_{\rm{0G}}$ and $\Omega_{\rm{0P}}$ from their scenarios as below
\begin{eqnarray}
\label{correlation-grishchuk}
\Omega_{\rm{0G}} \!&=&\! \frac{64\pi^3}{3} \times \left(\frac{l_{p}}{l_{H}}\right)^2 \left(\frac{\nu}{\nu_H}\right)^4 e^{2r_{\nu}} ~,\\
\label{correlation-parikh}
\Omega_{\rm{0P}} \!&=&\! \frac{32\pi^3 \nu^3}{3H_{0}^2} \times \Delta_h =\frac{64\pi^3 G\hbar \nu^4}{3H_0^2 c^5} \times   e^{2r_{\nu}} ~ .
\end{eqnarray}
The energy density $\Omega_{\rm{0G}}$ is derived with the strain given in \cite{grishchuk1} (Eq. 85). The energy density $\Omega_{\rm{0P}}$ is derived with the power spectrum  $\Delta_h=2G \hbar \nu/c^5 \times \cosh2r_{\nu}$ given in \cite{wilczek2} \footnote{Readers might notice that the amplification $\Delta_h$ is given by $e^{r_{\nu}}$ in \cite{wilczek2}, while in this work, the it is found to be $e^{2r_{\nu}}$.}, while the relation between the energy density and the strain power spectrum $\Omega_{\rm{0P}} = \frac{32\pi^3 \nu^3}{3H_0^2} \times \Delta_h$ is found from \cite{neil}. 



With the natural units, we have $l_p^2=G$, $l_H\nu_H=1$ and $\nu_H=H_0$. While plugging these expressions into (\ref{correlation-grishchuk}) and (\ref{correlation-parikh}), we are  surprised to find $\Omega_{0} = \Omega_{\rm{0P}} = \Omega_{\rm{0G}}$. The quantities of quantum noise from the three paradigms are the same.

\begin{figure}[h]
\centering
\includegraphics[width=80mm]{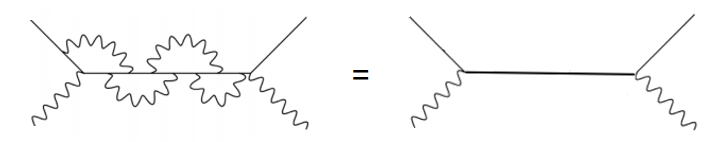}
\caption{\label{fig:interaction} Left: the Feynman diagram represents an elementary process of the full interaction between gravitons and the detector. Right: the leading order of the interaction. Solid lines represent the detector while wiggly ones represent gravitons. The symbol ``=''means the two Feynman diagrams contribute equally to the quantum noise. }
\end{figure}

It is easy to see that the quantum noise has the same origin: the quantization of gravity. This is addressed in \cite{wilczek1}. However, as Grishchuk et. al. just consider the leading order of the interaction, we can conclude that the quantum noise should completely come from that part. In another word, the higher orders of the interaction between gravitons and the detector, as well as the self-interaction, loops for both detector and gravitons, should contribute nothing to the quantum noise. Thus, with all these considerations, we illustrate this in Fig.\ref{fig:interaction}.

\section{The squeezed factor} 
All formulas given above are built in Minkowski space, and the squeezed factor, which describes the amplification of the gravitational wave, is dependent on the expansion of the universe. In co-moving space time, the squeezed factor is given as 
\begin{eqnarray}
e^{r_{\nu}(\eta)} \approx \frac{a(\eta)}{a(\eta_i)}
\end{eqnarray}
where $a(\eta)$ is the cosmological scale factor and $a(\eta_0)$ is the value of $a(\eta)$ at $\eta_i$, the beginning time of one expansion stage.

In \cite{grishchuk1}, the expansion of the universe is divided with four stages : i-stage is the inflation era with $a(\eta) \sim \eta^{1+\beta}$, z-stage is the ``stiffer''-dominated era with $a(\eta) \sim \eta^{1+\beta_s}$, e-stage is the radiation-dominated era with $a(\eta) \sim \eta$, and m-stage is the matter-dominated era with $a(\eta) \sim \eta^{2}$. In general, the index $\beta$ should be smaller than 0, and $\beta_s$ should be no larger than 0.  What we want to point out is: the stage paradigm is quite general, and the inflation can be included as one candidate of i-stage.

The amplitudes of gravitational waves, or graviton, with different wavenumbers are amplified at different stages, as is discussed in  \cite{grishchuk1}. Since the mode function of ours is the same as theirs (see \ref{eq:squeezed}), the squeezed factor they generated can be directly used here in principle, however, a new boundary of $\beta$ is found in this work upon our investigation.

\begin{figure}[h]
\centering
\label{stage}
\includegraphics[width=90mm]{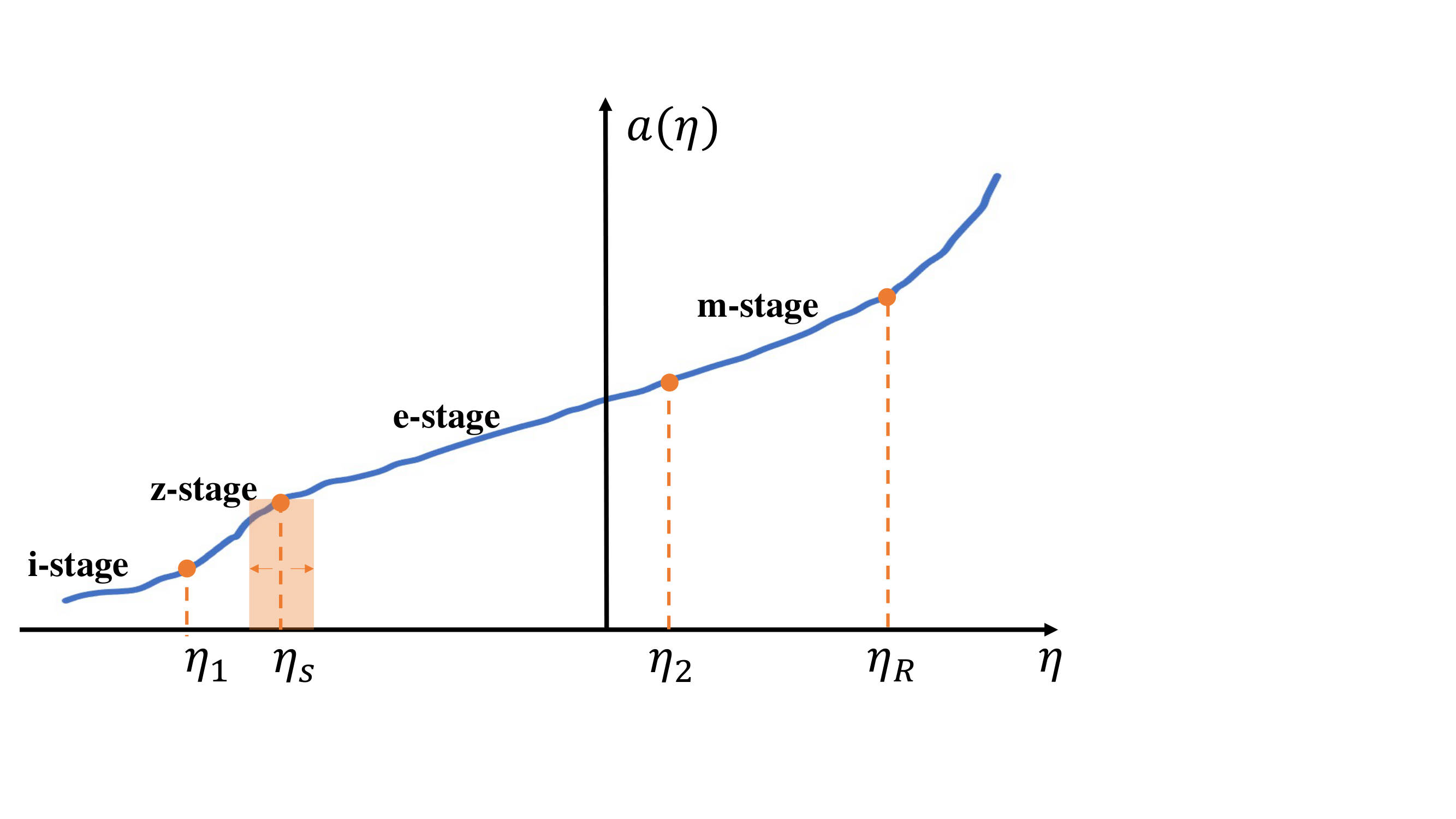}
\caption{\label{stage} Stages that our universe experienced. }
\end{figure}

The stages are shown in Fig.\ref{stage}, with the time at the end of each stage is $\eta_1$, $\eta_s$, $\eta_2$ and $\eta_R$ (the present time), while the corresponding wavenumbers are denoted as  $n_1, n_s$, $n_2$ and $n_H$ respectively. In this work, we use the description of laboratory frequencies $\nu_1, \nu_s$, $\nu_2$ and $\nu_H$, that are converted from $\nu= kc/2\pi a(\eta_R)$. Now let's estimate the squeeze factor below.

\subsection{The boundaries of the stages}

It is obvious that the squeeze factor is dependent on the index of  $\beta$ and $\beta_s$, and it is also dependent on the boundaries of different stages, $\nu_1$,$\nu_2$, $\nu_s$ and $\nu_H$, as the amplification is an integrated process. Grishchuk gave the following parameters with proper consideration in \cite{grishchuk1}  
\begin{equation}
\label{constraint-1}
\nu_1\!=\!3\times\!10^{10}Hz, \nu_2\!=\!2.3\times10^{-16}Hz,\nu_H\!=\!2.3\times10^{-18}Hz,
\end{equation}
where $\nu_H$ is the Hubble frequency, and $\nu_1$ is the highest frequency that the free graviton might not affect the rate of the primordial nucleosynthesis.  These parameters are taken directly in this work, while we discuss how to due with $\nu_s$.

In fact, the value of $\nu_s$, or how long the z-stage lasts is obscure. It is believed that the z-stage should include the process of baryogenesis, Pecci-Quinn symmetry breaking, the formation of cosmic topological defects. In this work, we consider a general case that z-stage can  simply be a part of the radiation-dominated era, with the lower frequency limit to be $\nu_s = 10^{-4}Hz$, while the upper limit, we set it to be $\nu_s = 10^{8}Hz$. Thus, the measurable range is included in both e-stage and z-stage.

\subsection{The boudnaries of $\beta$}

The consideration that the energy density of gravitational wave should be smaller than CMB leads us to
\begin{equation}
\label{constraint-2}
(\frac{\nu_1}{\nu_H})^{2+\beta} \leq 10^{6} \Rightarrow ~ \beta \leq -1.79 .
\end{equation}
Grishchuk observed an important equality as below \cite{grishchuk1}
\begin{eqnarray}
\label{eq:equality}
10^{-6} \frac{l_{H}}{l_{P}}  = (\frac{\nu_1}{\nu_H})^{-\beta} (\frac{\nu_1}{\nu_s})^{\beta_s} \frac{\nu_2}{\nu_H}.
\end{eqnarray}
Plugging the Planck length $l_P = 1.6\times 10^{-35} m$ and the Hubble length $l_H=1.3\times 10^{26} m$ into (\ref{eq:equality}), we obtain
\begin{eqnarray}
\label{constraint-3}
10^{53} (\frac{\nu_H}{\nu_1})^{-\beta} \!=\!(\frac{\nu_s}{\nu_1})^{-\beta_s},
\end{eqnarray}
so  $\beta_s$ can be written in $\beta$ as
\begin{eqnarray}
\label{constraint-33}
\beta_s = \frac{ln(10^{53}) \!+\!\beta ln(\nu_1)\!-\! ln(\nu_H)}{ln(\nu_1) \!-\!ln(\nu_s)}.
\end{eqnarray}
Thus, we can use one parameter $\beta$ for the squeezed factor. Considering $\beta_s \leq 0$, a new constraint using (\ref{constraint-3}) is found 
\begin{eqnarray}
\label{constraint-4}
\beta \leq \frac{ln(10^{53})}{ln\nu_{H} -ln \nu_1}=-1.89 ~.
\end{eqnarray}
Further, if we ask the integral (\ref{eq:noise}) to be convergent, a lower limit of  $\beta \geq -2$ is obtained, and thus, we obtain a full constraint for $\beta$  
\begin{equation}
\label{constraint-full}
-2.0 \leq\beta \leq -1.89 ,
\end{equation}
where the case with $\beta=2.0$ is known as the Harrison-Zeldovich spectrum.

\begin{figure*}[ht]
\centering
\includegraphics[width=120mm]{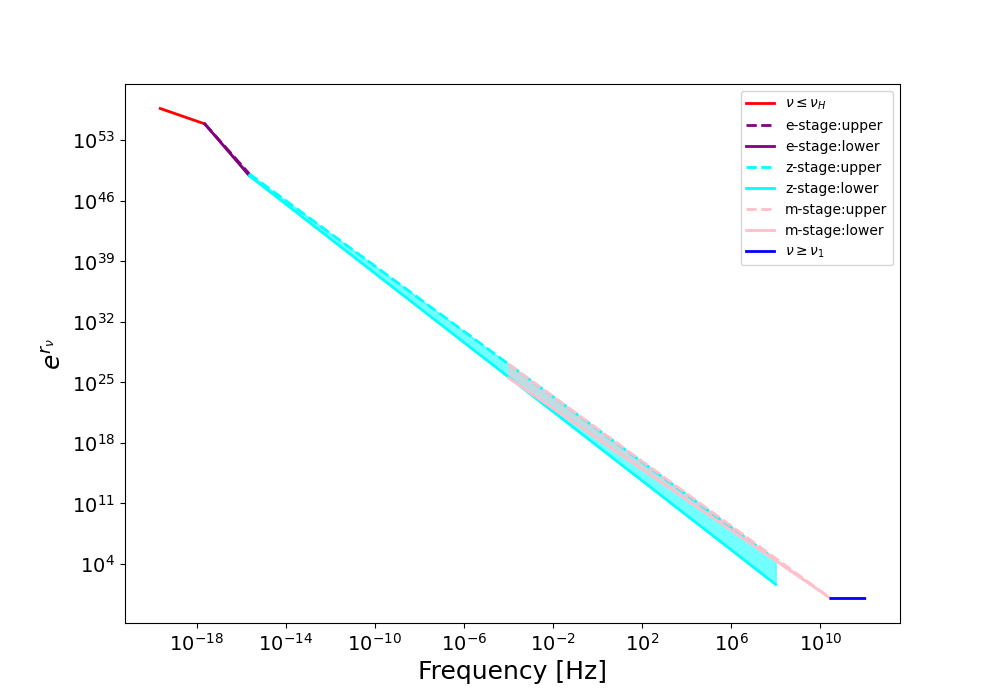}
\caption{\label{squeezed}The squeezed factor for different stages. }
\end{figure*}

\subsection{The boundary of the squeezed factor}
We have pointed out that the squeezed factor is only dependent on the parameter $\beta$, and thus, if it is a monotonical function, the boundaries will be determined by $\beta_{min}$ and $\beta_{max}$. This is discussed below, based on the squeezed factor given in \cite{grishchuk1} for each stage.

In e-stage, the squeezed factor is given as 
\begin{eqnarray}
\label{eq:e-stage} 
e^{r_\nu} = \left(\frac{\nu}{\nu_s}\right)^{\beta}
\left(\frac{\nu_s}{\nu_1}\right)^{\beta - \beta_s}\!,~(\nu_2 < \nu < \nu_s)~,
\end{eqnarray}
with the help of formula (\ref{constraint-3}), we have
\begin{eqnarray}
\label{eq:e-stage-2} 
e^{r_\nu} = (\frac{\nu}{\nu_H})^{\beta}\times 10^{53} .
\end{eqnarray}
We see the squeezed factor $e^{r_{\nu}}$ monotonically decreases with $\beta$. Two groups of proper parameters satisfying (\ref{constraint-full}), with a maximal $\beta_{max}$ and a minimal $\beta_{min}$ are 
\begin{eqnarray}
\label{parameters-1_1}
\nu_s \!&=&\! 3\times 10^{8}Hz, ~~\beta_{max}\! =\! -1.89, ~~ \beta_s \!=\! -0.12,\\
\label{parameters-1_2}
\nu_s \!&=&\! 3\times 10^{8}Hz, ~~\beta_{min} \!=\! -2.0, ~~ \beta_s \! =\! -1.66. 
\end{eqnarray}
In z-stage, the squeezed factor is given as
\begin{eqnarray}
\label{eq:z-estage}
e^{r_\nu} = \left(\frac{\nu}{\nu_1}\right)^{\beta-\beta_s} ~,
~~\left(\nu_s < \nu < \nu_1\right),
\end{eqnarray}
where the index $\beta -\beta_s$ is given with the form of
\begin{eqnarray}
 \beta-\beta_s = \frac{1}{2}a_0 \beta + \frac{1}{2}b_0 
\end{eqnarray}
where $a_0= \frac{ln(v_{s}/v_{H})}{ln(v_s/v_1)}=-0.94$ and $b_0= \frac{ln(10^{53})}{ln(v_s/v_1)}=-3.66$.  It is clear that  $\beta -\beta_s$  monotonically decreases with $\beta_s$, while  $e^{r_{\nu}}$ monotonically increases with $\beta_s$. Two groups of parameters with a maximal $\beta_{max}$ and a minimal $\beta_{min}$ satisfying (\ref{constraint-full}) are 
\begin{eqnarray}
\label{parameters-2_1}
\nu_s \!&=&\! 10^{-4} Hz,~~ \beta_{max} \!=\! -1.89, ~~ \beta_s \!=\! -0.017,\\
\label{parameters-2_2}
\nu_s \!&=&\! 10^{-4} Hz,~~ \beta_{min} \!=\! -2.0, ~~ \beta_s \!=\! -0.24. \end{eqnarray}
In m-stage, the squeezed factor is given as
\begin{eqnarray}
\label{eq:z-estage}
e^{r_\nu} = \left(\frac{\nu}{\nu_2}\right)^{\beta-1}\!\! \left(\frac{\nu_2}{\nu_1}\right)^{\beta} \left(\frac{\nu_s}{\nu_1}\right)^{-\beta_s}, ~~ \nu_H \leq \nu \leq \nu_2 ~,
\end{eqnarray}
and with (\ref{constraint-33}), the expression becomes
\begin{eqnarray}
e^{r_\nu} = (\frac{\nu}{\nu_H})^{\beta -1} \frac{\nu_2}{\nu_H} \times 10^{53} .
\end{eqnarray}
We find $e^{r_{\nu}}$ monotonically decreases with $\beta$. However, except the m-stage, e-stage and z-stage, we have two other possibilities, which are shown below
\begin{eqnarray}
e^{r_{\nu}} \!&=&\! (\frac{\nu}{\nu_H})^{\beta +1}(\frac{\nu_H}{\nu_2})^{\beta-1}(\frac{\nu_2}{\nu_1})^{\beta}(\frac{\nu_s}{\nu_1})^{-\beta_s}, ~~\nu \leq \nu_H, \nonumber \\
e^{r_{\nu}} \!&=&\!1, ~~ \nu \geq \nu_1 .
\end{eqnarray}
The squeezed factor with different stages is shown in Fig. {\ref{squeezed}}. The upper boundaries and the lower boundaries for m-stage, e-stage and z-stage are illustrated. The overlapping range of e-stage and z-stage is due to the obscure of $\nu_s$ that has been discussed above. We set the lower frequency limit of the measurement to be $10^{-20} Hz$, while the upper frequency limit is set to be $10^{12} Hz$.

\begin{figure*}[ht]
\centering
\includegraphics[width=120mm]{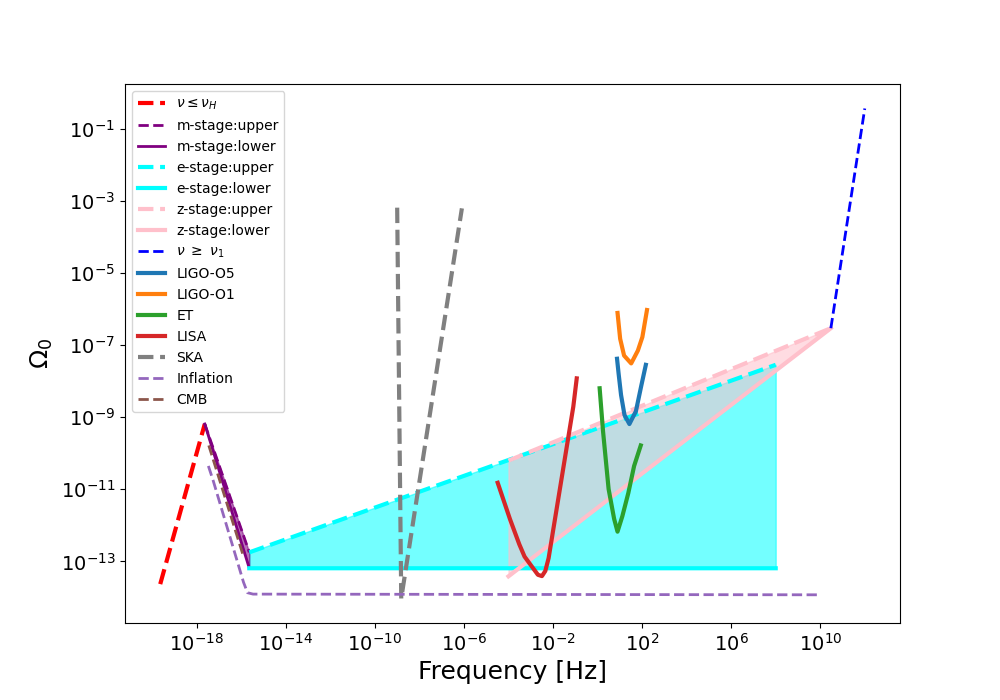}
\caption{\label{spectrum}The detectable area with the present-day spectrum $\Omega_{0}$ from quantum noise induced by gravitons. The filled cyan region represents the detectable area of quantum noise from the e-stage. The pink region represents the detectable area from the z-stage, while the purple region is the detectable field from m-stage. The red dash line is the spectrum for $\nu \leq \nu_H$, while the other side, the spectrum for $\nu \geq \nu_1$ is illustrated. The sensitivities of space-based and ground-based detectors are illustrated. 'ET' is the third generation detector Einstein Telescope. The spectrum of inflation from slow-roll is shown. The sensitivity of SKA with grey dash line is presented for discussion. }
\end{figure*}

\section{The Detectability} 

It is observed that, the spectrum of the energy density is in an unified form for each stage and can be described by  
\begin{equation}
\label{unify}
\Omega_0 \sim \nu^{4+ a_0 \beta+ b_0} ~,
\end{equation}
where $a_0 \beta+ b_0$ comes from the contribution of the squeezed factor. For z-stage, $a_0=-0.94$ and $b_0=-3.66$ as we have shown above, while for e-stage, we can take $a_0 =2$ and $b_0 =0$. For m-stage, we use  $a_0 =2$ and $b_0 =-2$.\\

The spectrum from (\ref{eq:omega}) is illustrated in Fig. \ref{spectrum}. The cyan area in the figure is the detectable field from e-stage, and the upper and lower boundaries are generated from the parameters shown in
(\ref{parameters-1_1}) and (\ref{parameters-1_2}) respectively. The lower boundary is flat, and it is corresponding to the Harrison-Zeldovich spectrum with $\Omega \sim \nu^0$. The pink area in the figure is from z-stage, with the boundaries generated from parameters in (\ref{parameters-2_1}) and (\ref{parameters-2_2}), while the purple area is the detectable filed from m-stage.
The overlapping range of e-stage and z-stage comes from the obscureness of the $\nu_s$, as we don't know how long z-stage lasts.  Thus, it is fair to say that, in case such a spectrum is found, the quantum noise from gravitions  might be confirmed. We would like to give some clarifications below.

From Fig. \ref{spectrum}, we see the quantum noise might be detected by both the ground-based detectors like LIGO-O5, ET, and the space-based detectors like LISA, as their sensitivity curves are inside of the detectable area. But for detector LIGO-O1, the quantum noise is difficult to detect as its sensitivity is out of the detectable area.


Gravitational waves from all sources produce quantum noise if they arrive at the detectors. However, only the gravitational waves background from the early universe in squeezed states, could produce enough energy density and are able to be detected. This is because the amplitude of the graviton is amplified during the evolution of the universe with an exponential form. It is highly possible that the process of making the gravitons into quantum state is the inflation \cite{grishchuk1,albrecht}. 

The quantum noise in non-squeezed state is much smaller than that from the classical background, as it is the perturbative quantity of the gravitational waves background. This can be confirmed by setting $e^{r_{\nu}} =1$ in (\ref{eq:omega}) for all the stages, where the spectrum in Minkowski state is generated. At $\nu = 0.1 Hz$, $\Omega_0$ is found to be around $10^{-54}$, while at $\nu = 10 Hz$, $\Omega_0$ is found to be around $10^{-46}$. It is clear to see that the energy density is quite small, far beyond the detectability of any current or upcoming detector.

As we have pointed out, the non-stationary feature of the signal is distinguishing in principle. However, the study of \cite{allen} indicates, this looks hopeless because the feature is statistical which needs a measurement as long as the universe age. In this work, we show that the spectrum from quantum noise at each stage might be some unique. In the spectrum $\Omega_0 \sim \nu^{4+ a_0 \beta+ b_0}$,  $a_0$ and $b_0$ in the are constant for each stage, while $\beta$ is a variable. It is easy to see that: for e-stage, $0 \leq 4+ a_0 \beta+ b_0 \leq 0.22$, for z-stage, $2.12 \leq 4+ a_0 \beta+ b_0 \leq 2.22$, while for m-stage, $-2 \leq 4+ a_0 \beta+ b_0 \leq -1.78$.

The calculation of the correlation in squeezed state is also shown in \cite{soda}, where the authors use the effective stain, which has the same physical meaning as the quantum noise in this letter. However, their calculation of the correlation looks indirect and obscure to us while a different spectrum with the form of $\Omega \sim \nu^{3+ a_0 \beta+ b_0}$ will be found, if the squeezed factor for each stage has been considered.

As shown in the figure, the minimal point of the sensitivity of SKA is obviously below the detectable area of the quantum noise. However, the quantum noise can not be detected through indirect detection, as there is no direct interaction between gravitons and the detectors. A similar is for quantum noise below $\nu_2$ in the detectable range of CMB: thought experiments like SP4, BICEPT, Planck are designed to search for gravitational waves background, they cannot be used to detect the quantum noise.

\section{Summary} 

To summarize, quantum gravity is what physicists are chasing yet to be completed. Some physicists think that gravity may not be quantized canonically, and if so, the existence of gravitons is questionable. We studied the detectability of quantum noise in the squeezed state from the induced gravitons in this work. 

It is surprise to observe that our result is the same as what Grishchuk at.al and Parikh et.al. gave in the previous work. We believe this implies that our three groups are calculating the same thing: the correlation of the quantum noise from the leading order of the interaction between the graviton and the detector. 

With the squeezed factor properly estimated, a spectrum of the form $\Omega_0 \sim \nu^{4+a_0 \beta+b_0}$ is found, and possible detectable ranges were given with the available $\beta$. Compared with the sensitivities of the current and upcoming detectors, a conclusion that the quantum noise might be detected from the upcoming detectors such as LIGO-O5, LISA and ET, was made. 

\section{Acknowledgements} 
X. F. is supported by the National Natural Science Foundation of China(under Grants No.11922303) and Hubei province Natural Science Fund for the Distinguished Young Scholars (2019CFA052).\\


\begin{thebibliography}{}
\bibitem{jacobson} T.~Jacobson, Phys. Rev. Lett. \textbf{75}, 1260-1263 (1995).

\bibitem{feyman1} R. P. Feynman,  Acta Phys. Polon. 24, 697 (1963).

\bibitem{feyman2} S. Weinberg, Phys. Lett. 9, 357 (1964).

\bibitem{feyman3} S. Deser, Gen. Rel. Grav. 1, 9 (1970).

\bibitem{feyman4} D. G. Boulware and S. Deser, Annals Phys. 89, 193 (1975).

\bibitem{dyson}F.~Dyson, Int. J. Mod. Phys. A \textbf{28}, 1330041 (2013).

\bibitem{Kanno2018} S.~Kanno and J.~Soda, Phys. Rev. D \textbf{99}, no.8, 084010 (2019).

\bibitem{Kanno2019} S. Kanno, Phys. Rev. D 100, no.12, 123536 (2019).

\bibitem{allen} B.~Allen, E.~E. Flanagan and M.~A. Papa, Phys.Rev. D 61, 024024 (2000).

\bibitem{grishchuk1} L.~P. Grishchuk, Phys.Usp. 44 (2001) 1-51; Usp.Fiz.Nauk 171 (2001) 3-59. 

\bibitem{grishchuk2} L.P. Grishchuk, Lect.Notes Phys. 562 (2001) 167-194.

\bibitem{grishchuk3} L.P. Grishchuk, Phys.Usp. 48 (2005). 


\bibitem{vedral1}
S.~Bose, A.~Mazumdar, G.~W.~Morley, H.~Ulbricht, M.~Toroš, M.~Paternostro, A.~Geraci, P.~Barker, M.~S.~Kim and G.~Milburn, Phys. Rev. Lett. \textbf{119}, no.24, 240401 (2017). 

\bibitem{vedral2}
C.~Marletto and V.~Vedral,  Phys. Rev. Lett. \textbf{119}, no.24, 240402 (2017).

\bibitem{vedral3}
A.~Bassi, A.~Großardt and H.~Ulbricht, Gravitational Decoherence, Class. Quant. Grav. \textbf{34}, no.19, 193002 (2017).

\bibitem{wilczek1} M.~ Parikh, F.~ Wilczek, G.~ Zahariade, Int.J.Mod.Phys.D 29 (2020) 14, 2042001.

\bibitem{wilczek2} M.~ Parikh, F.~ Wilczek, G.~ Zahariade, e-Print: 2010.08205.
\bibitem{wilczek3} M.~ Parikh, F.~ Wilczek, G.~ Zahariade, e-Print: 2010.08208.

\bibitem{hu1} C. Anastopoulos, Phys. Rev. D 54, 1600-1605 (1996).

\bibitem{hu2} M. P. Blencowe, Phys. Rev. Lett. 111, no.2, 021302 (2013).

\bibitem{hu3} C. Anastopoulos and B. L. Hu, Class. Quant. Grav. 30, 165007 (2013).

\bibitem{hu4} Breuer, H.-P., Petruccione, Phys. Rev. A 63, 032102 (2001).

\bibitem{hu5} C. J. Riedel, arXiv:1310.6347.

\bibitem{hu6} T. Oniga and C. H. T. Wang, Phys. Rev. D 93, no.4, 044027 (2016).

\bibitem{hu7} T. Oniga and C. H. T. Wang, Phys. Rev. D 96, no.8, 084014 (2017) .

\bibitem{hu8}J. Gamboa, R. MacKenzie, F. Mendez, e-Print: 2010.12966.

\bibitem{zurek} K.~M.~Zurek, e-Print: 2012.05870. 

\bibitem{soda} S.~ Kanno, J.~ Soda, J.~ Tokuda, Phys. Rev. D 103, 044017 (2021).

\bibitem{neil} J.~D.~Romano, N.~J.~Cornish, Living Rev.Rel. 20 (2017) 1, 2. 


\bibitem{albrecht} A.~Albrecht, P.~Ferreira, M.~Joyce, and T.~Prokopec, Phys. Rev. D 50, 4807 (1994); 








\end{thebibliography}
\end{document}